\documentclass[11pt]{article}

\usepackage{graphicx}
\usepackage[font={footnotesize,stretch=1}, labelfont=bf]{caption}
\usepackage[letterpaper,margin=1in]{geometry}
\usepackage[colorlinks=true,linkcolor=blue,	filecolor=blue,	citecolor=blue,urlcolor=blue]{hyperref}
\usepackage[backend=biber,style=nature,url=false,doi=false,isbn=false,date=year]{biblatex}
\renewenvironment{abstract}
	{\quotation}
	{\endquotation}

\date{}

\makeatletter
\renewcommand{\fnum@figure}{\textbf{Figure \thefigure}}
\renewcommand{\fnum@table}{\textbf{Table \thetable}}
\makeatother

\usepackage{url}
\usepackage[normalem]{ulem}

\def\scititle{
	Density and curvature govern the dynamical self-organization of active matter on a closed surface
}

\title{\bfseries \boldmath \scititle}

\author{
    \parbox[t]{0.95\textwidth}{
        \centering
        Romain~Leroux$^{1}$,
        Violeta~Zilia~Marcen~Arroyo$^{2}$,
        André~Estevez-Torres$^{3}$,\\
        Andrew~Callan-Jones$^{2}$,
        Raphael~Voituriez$^{1}$,
        Ananyo~Maitra$^{1,4}$,\\
        Nicolas~Lobato-Dauzier$^{1\ast}$,
        Jean-Christophe~Galas$^{1\ast}$\\[6pt]
        \smallskip
        \footnotesize
        $^{1}$Sorbonne Université, CNRS, Institut de Biologie Paris-Seine (IBPS), Laboratoire Jean Perrin (LJP), F-75005, Paris, France.\\[6pt]
        $^{2}$Laboratoire Matière et Systèmes Complexes, Université Paris Cité/CNRS (UMR 7057), Paris, France.\\[6pt]
        $^3$Université de Lille, CNRS, Laboratoire de spectroscopie pour les interactions, la réactivité et l’environnement (UMR 8516 - LASIRE), F-59655, Villeneuve d’Ascq, France.\\[6pt]
        $^4$CY Cergy Paris Université, CNRS, Laboratoire de Physique Théorique et Modélisation, 95032 Cergy-Pontoise, France.\\[6pt]
        $^\ast$\textbf{Corresponding authors:} Nicolas Lobato-Dauzier, Jean-Christophe Galas.\\
        Email: \{nicolas.lobato-dauzier; jean-christophe.galas\}@sorbonne-universite.fr\\[6pt]
        \textbf{Author Contributions:} RL, AET, NLD, JCG designed the research. RL performed the experiments. RL, NLD, JCG analyzed the data. VZMA, ACJ, RV, AM performed the analytical modelling. NLD, JCG supervised the research. RL, VZMA, AET, ACJ, RV, AM, NLD, JCG wrote the paper.\\[6pt]
        \textbf{This PDF file includes:} Main Text, Figures 1 to 6.
    }
}

\begin{document} 

\maketitle

\begin{abstract} \bfseries \boldmath
In active and living matter, topological defects provide a powerful fingerprint, reducing complex nematic organization to a small number of characteristic structures that can localize stresses, reorganize tissues and define morphological axes, all key features of morphogenesis. Understanding how topological defects organize and evolve in the complex geometries and heterogeneous distributions of active material found in biological systems is therefore essential.
Here, we vary the geometry and density of a minimal in vitro self-organizing microtubule–kinesin cortex formed at the surface of water-in-oil droplets. We find that the active material reorganizes through topological transitions, with the spontaneous emergence of additional positive and negative defects and the selection of dynamical deterministic patterns.

We rationalise our findings via an active matter-based theoretical framework that incorporates a generic coupling of the dynamics to curvature. The intrinsic coupling between nematic orientation and polar flow, together with topological and geometrical constraints, allows us to identify selection rules governing dynamical states of increasing complexity. Remarkably, this minimal framework captures the complexity of the experimentally observed dynamics. Our results show that deterministic morphogenetic-like cycles, defect topology, and material organisation may arise from the interplay of activity, density, and curvature.

\end{abstract}

\subsection*{Introduction}

Can geometry and mechanics regulate complex morphogenetic dynamics?  Morphogenesis, the emergence of form in biological systems, results from an interplay between geometry, mechanics, and biochemical signaling \cite{collinet_programmed_2021}. These components act as intertwined sources of morphogenetic information: geometry dictates stresses and flows, mechanics reorganizes material and alters geometry, and biochemical signaling interprets and feeds back onto these cues. These mutual couplings make it difficult to determine which morphogenetic features require biochemical regulation and which can arise from physical interactions alone. Living systems further complicate this challenge, as perturbations often activate compensatory pathways that obscure direct mechanisms \cite{liu_morphogenesis_2024}. By contrast, in vitro synthetic active materials provide a framework to control these factors while retaining nonequilibrium activity \cite{Sciortino2025}. Here we couple active matter and spherical confinement and show how simple geometric ingredients such as surface density and curvature can generate robust dynamic self-organization.

In recent years, microtubule–kinesin active gels, derived from the cellular cytoskeleton, have emerged as powerful model systems for active matter self-organization \cite{nedelec_self-organization_1997,senoussi_tunable_2019,shankar_topological_2022,senoussi_tunable_2019}. These reconstituted materials display spontaneous motion and collective behaviors including large-scale contractions \cite{bendix_quantitative_2008,foster_active_2015}, chaotic flows \cite{sanchez_spontaneous_2012}, defect-mediated patterns \cite{Guillamat2018}, and coherent streaming states \cite{wu_transition_2017,Sakuta2023}. In active nematics, topological defects not only structure the director field, but, due to activity, also act as local sources of motion, behaving as self-propelled particles \cite{marchetti_hydrodynamics_2013}. Beyond individual defects, confinement and boundary conditions strongly influence large-scale patterns and flows in active nematics, selecting coherent circulating states, defect ordering, and system-spanning vortices \cite{alam_active_2024,Hardouin2019}. While such phenomena have been extensively characterized in planar or weakly curved geometries, the role of strongly curved confinement, ubiquitous in cellular and developmental contexts, has only recently begun to be explored experimentally \cite{alam_active_2024, keber_topology_2014, hsu_activity-induced_2022, ellis_curvature-induced_2018}.

Biologically relevant in the context of cell cortices and embryonic tissues, spherical confinement is particularly interesting because it combines curvature and topological constraints: a closed surface enforces a total topological charge of $+2$, couples active stresses to curvature, and can channel activity into persistent, symmetry-breaking dynamics. On a sphere, the minimal equilibrium defect configuration consists of four +1/2 nematic defects arranged such that they form the vertices of a regular tetrahedron, with defects repelling each other to maximize their distance. 
Complementing several theoretical studies \cite{nestler_active_2022, henkes_dynamical_2018, zhang_dynamics_2020, vetter_morphogenesis_2014,alaimo_curvature_2017}, a seminal experiment by Keber \textit{et al}. \cite{keber_topology_2014} demonstrated that four self-propelled +1/2 defects on a spherical active nematic vesicle oscillate between tetrahedral and planar configurations. However, morphogenetic active materials rarely correspond to uniformly coated surfaces of constant curvature: cytoskeletal and cortical filaments are often heterogeneously distributed, and closed cellular or tissue-scale surfaces can display strong spatial variations in curvature. How do geometry and the spatial distribution of active material constrain defect organization and give rise to complex, yet deterministic, collective dynamics?

Here, by encapsulating a microtubule–kinesin active gel inside droplets, we realize a system in which confinement, surface density, and curvature can be independently tuned. Microtubules accumulate at the oil–water interface and self-organize into a two-dimensional active nematic cortex hosting four +1/2 topological defects. By varying the surface density, we show that the system continuously spans distinct dynamical regimes, from defect-dominated nematic dynamics to a filament-like regime characterized by continuous reorganization, while geometric deformation through squeezing triggers defect nucleation and topological restructuring. We further establish simple, experimentally validated rules that combine nematic alignment with geometry and topology to infer compatible flow architectures.

This framework may help turn static observations of nematic organization in biological systems into plausible dynamical scenarios, offering insight into how biological architecture and shape govern collective dynamics.

\newpage
\subsection*{Experimental system and qualitative results}

\begin{figure}[htbp]
    \centering
    \includegraphics[width=0.95\textwidth]{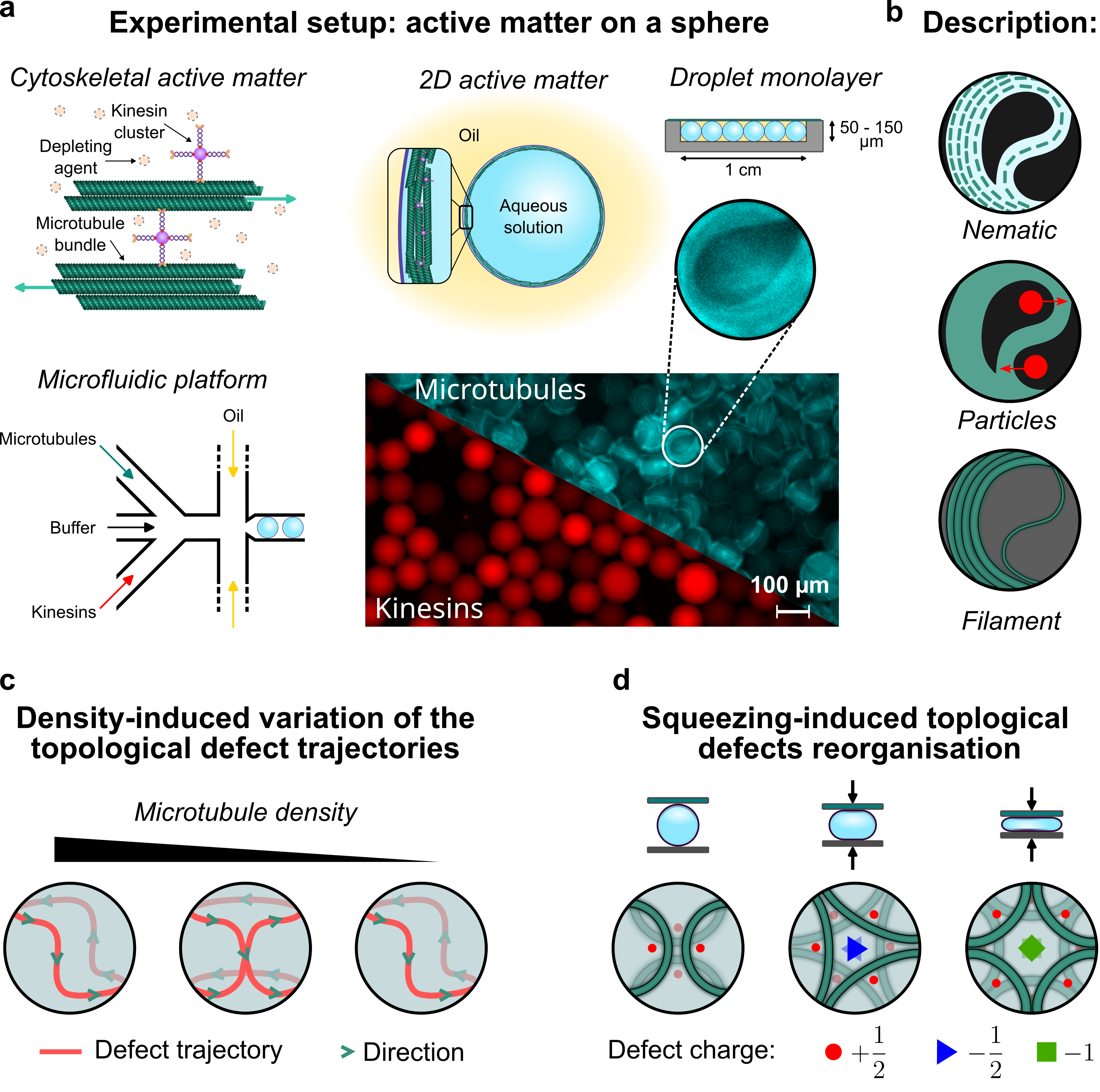}
    \caption{\textbf{Experimental system and emergent active nematic organization.} \textbf{(a)} Schematic of the experimental setup. Top left: The active matter is formed from non-growing microtubules bundled by a depleting agent and driven by multimeric kinesin motor clusters. Bundles are on average apolar allowing kinesins to walk toward both directions, generating extensile sliding between the former. Bottom left: Three-inlet microfluidic device for generating water-in-oil droplets with controlled motor and filament concentrations. Top right: Cross-sectional view of a droplet showing microtubules (green) accumulating at the oil–water interface to form a two-dimensional active nematic cortex. The droplets are then placed in a silicon chamber forming a monolayer, allowing the imaging. Bottom right: Fluorescence microscopy image showing a layer of droplets with varying kinesin (red) and microtubule (cyan) concentrations. An isolated droplet shows the phase separation between microtubule-rich regions (bright) and microtubule-poor voids (dark empty area). \textbf{(b)} Schematic illustrating multiple complementary descriptions of the active cortex: as a continuous active nematic field (top), as interacting topological defects (center, red dots mark +1/2 defects), or as an extensile filament structure (bottom). \textbf{(c)} Density-dependent trajectories of defects: the coiling extent varies non-monotonically with density. \textbf{(d)} Geometry-dependent reorganization. Spherical droplets (left) host four +1/2 defects. Squeezing (center, right) creates flat regions where +1/2 defects cannot reside, confining them to curved peripheral zones. It also triggers nucleation of -1/2 or –1 defects, maintaining a total topological charge of +2 while adapting to geometric constraints.}
    \label{figure_1}
\end{figure}

The active material consists of stabilized microtubules bundled by a depletion agent and driven by multimeric kinesin clusters (Fig. \ref{figure_1}a, top left). Because bundles are, on average, apolar, kinesin motors do not produce ballistic transport but generate extensile sliding between adjacent filaments. ATP regeneration maintains activity over hours, allowing sustained active stresses throughout the duration of the experiments.

Water-in-oil droplets were produced by microfluidics, allowing independent control of droplet size, motor concentration, and filament concentration (Fig. \ref{figure_1}a, bottom left, Fig. S1 and S2, Movie S1). After encapsulation, active microtubule bundles accumulate at the oil–water interface and form a two-dimensional cortex over tens of minutes (Fig. S4). Fluorescence imaging reveals a dynamic cortex composed of microtubule-rich aligned regions and microtubule-poor voids (Fig. \ref{figure_1}a, right). Imaging both hemispheres enables reconstruction of defect trajectories on the full droplet surface.

At high microtubule density, the interfacial active layer is well described as a canonical active nematic, characterized by an orientation field that is well defined everywhere except at the location of defects, and four +1/2 defects embedded 
(Fig. \ref{figure_1}b, Fig. S3, Movie S2). Upon reducing the microtubule density, the system undergoes a qualitative change in appearance while retaining similar large-scale dynamics. The cortical pattern now resembles a single thick extensile filament that circulates and folds along the droplet interface, suggesting a transition to an alternative physical description based on filament mechanics rather than nematic hydrodynamics (Fig. \ref{figure_1}b).
Despite their distinct visual and conceptual interpretations, we will show that these two regimes represent complementary views of a continuous phenomenon (Fig. \ref{figure_1}c). The filamentous regime retains local nematic order with defect-like structures localized at bending points, while the nematic regime exhibits dense filamentous bands spanning between defects. In both cases, the system displays periodic coiling dynamics, although the extent of coiling varies non-monotonically with density. This indicates that a unified mechanism that couples mechanics and geometry governs organization across all densities, with density modulating the effective degrees of freedom available to the system.

Finally, when droplets are mechanically squeezed between two surfaces, curvature is redistributed and flat central regions with near-zero curvature emerge. In response, additional defects nucleate to accommodate the altered geometric constraints, demonstrating the sensitivity of defect organization to curvature heterogeneity (Fig. \ref{figure_1}d). 
This versatility establishes spherical droplets as a tractable platform for studying how activity, density, and curvature jointly govern active nematic dynamics and the coupling between mechanics and geometry.

\newpage
\subsection*{Periodic defect dynamics on the sphere: Coiling and crossing phases}

To quantify defect motion, we reconstructed the full droplet surface by simultaneously imaging both hemispheres and tracking the four +1/2 defects over time (Fig. \ref{figure_2}a and 2c, Movie S3). Defects were identified as regions of highest curvature in the nematic director field, coinciding with microtubule-depleted, small teardrop-shaped voids whose tips correspond to the points of maximal curvature.

From the reconstructed defect positions, we computed pairwise geodesic angles and spherical coordinates to quantify collective organization and individual motion (Fig. \ref{figure_2}d–f and Fig. S5 to S7). Missing equatorial segments were reconstructed using the symmetry and periodicity of the six pairwise distances, as detailed in the supplementary materials section 4.

Time traces reveal regular oscillations with a period of approximately 500 seconds between two characteristic configurations (Fig. \ref{figure_2}b): (1) a tetrahedral configuration, with all pairwise geodesic angles near 109.5$^\circ$ ($\sim$3$\pi$/5 radians), corresponding to vertices of a regular tetrahedron inscribed on the sphere, and (2) a planar configuration, with defects lying approximately on a great circle—two angles near 180$^\circ$ ($\pi$) and four near 90$^\circ$ ($\pi$/2).
The amplitude and regularity are remarkably robust across multiple droplets, indicating that periodic reorganization is an intrinsic property of the active spherical nematic.

The motion decomposes into two alternating phases: coiling within one hemisphere and crossing to the opposite hemisphere.

During the coiling phase, two pairs of defects circulate around the two poles in a coordinated manner. For each defect, the azimuthal angle $\varphi$ evolves rapidly while the polar angle $\theta$ remains nearly constant, indicating that defects stay at approximately the same latitude. Defects trace nearly circular paths with synchronized motion maintaining approximately equal angular spacing. 

The system reaches a tetrahedral arrangement as defects become more evenly distributed across hemispheres, but activity prevents stabilization: the tetrahedron collapses with defects that turn and move toward the equator and beyond.
During the crossing phase, defects collectively transition to the opposite hemisphere. Now $\theta$ changes rapidly as defects move from one pole toward the other, while $\varphi$ remains approximately constant - defects move along meridians. This behaviour typically lasts for a shorter period of time (50 seconds) than coiling (200 seconds) and corresponds to all four defects meeting at the equator before switching hemispheres. 
These phases repeat periodically, producing a self-sustained trajectory cycling between tetrahedral and planar states that persists for hours without damping. 
Mutual defect repulsion favors a tetrahedral arrangement that maximizes pairwise separations, but this configuration is destabilized by active stresses and defect polarity \cite{keber_topology_2014}. During the coiling phase, extensile forces drive defects toward a planar configuration, which in turn becomes unstable as continued activity pushes defects across the equator. Crossing redistributes defects between hemispheres, allowing the system to relax back toward a tetrahedral-like arrangement and thereby complete the cycle.

\begin{figure}[p]
    \centering
    \includegraphics[width=1\textwidth]{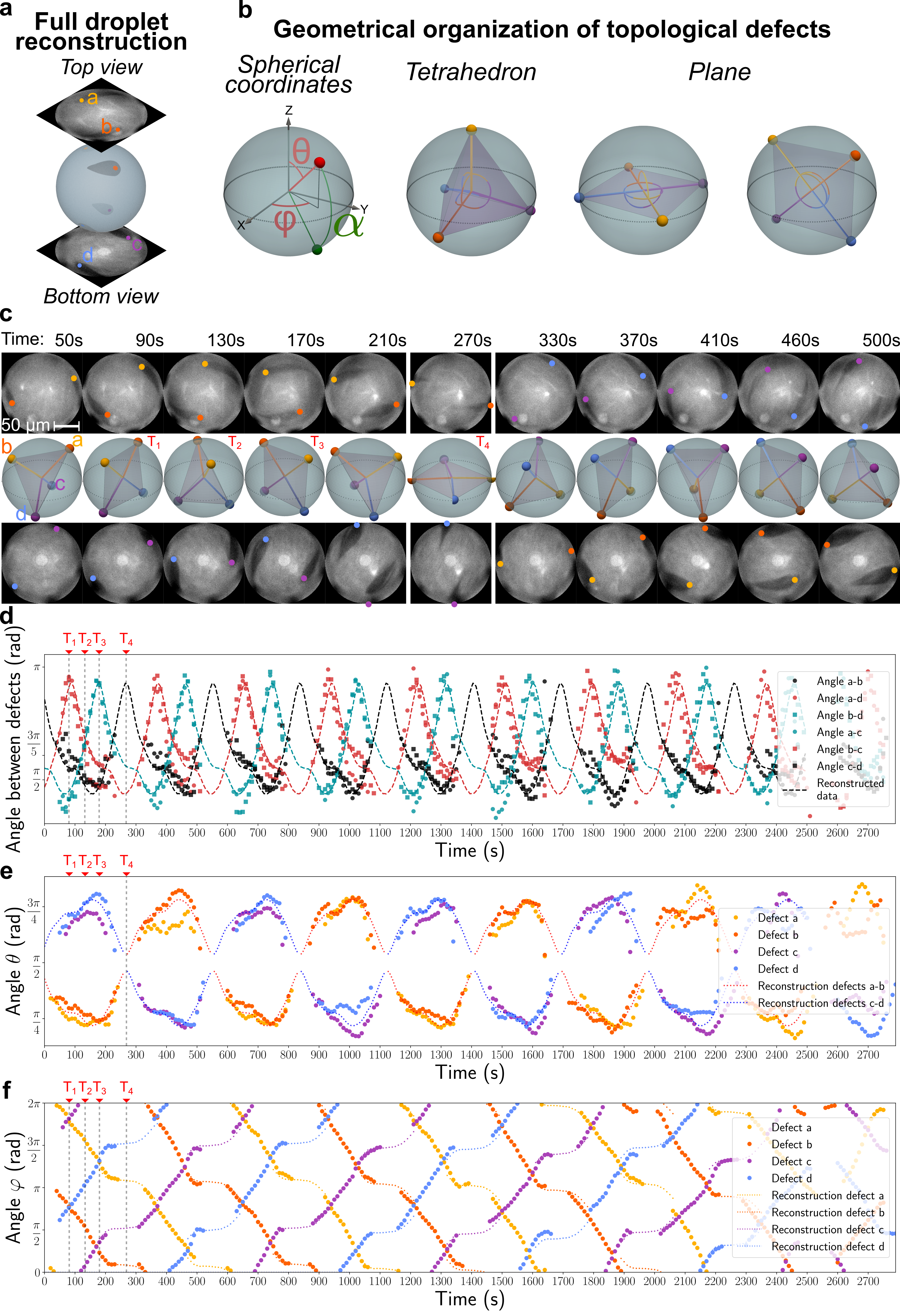}
\end{figure}

\clearpage
\captionof{figure}{\textbf{Periodic dynamics of topological defects on a nematic sphere.} \textbf{(a)} Three-dimensional reconstruction of the full droplet surface showing trajectories of the four +1/2 topological defects over time. Defects are labeled and color-coded (a: yellow, b: orange, c: purple, d: blue). Both hemispheres are reconstructed from simultaneous top and bottom imaging. \textbf{(b)} Schematics illustrating spherical coordinates and geodesic angles, with the two characteristic spatial configurations. Left: Tetrahedral configuration, with defects positioned at the vertices of a regular tetrahedron inscribed on the sphere, all pairwise angles near 109.5°. Right: Planar configuration, with defects arranged approximately on a great circle, two angles near 180° and four near 90°. \textbf{(c)} Timelapse images of the top and bottom hemispheres of the droplet with the identified defects over one period of defect movement. Middle: reconstructed positions of the defects on a 3D sphere, and visualization of the polygon they shape (pink). \textbf{(d)} Time series of the six pairwise geodesic angles between defects, color-coded by symmetry group. The angles oscillate periodically between the planar state (two angles $\sim$180°, four $\sim$90°) and the tetrahedral state (all six $\sim$109.5°). Period $\approx$ 500 seconds. Empty regions are linked to the defects not imaged when on the equator. Dashed lines are a reconstruction of the angles using symmetry rules. These reconstructions are also used for the following angles $\theta$ and $\varphi$. \textbf{(e)} Time series of the polar angle $\theta$ of the same defects, demonstrating the same coiling ($\theta$ $\approx$ constant) and changing hemisphere ($\theta$ varying) phases. \textbf{(f)} Time series of the azimuthal angle $\varphi$ of the four defects showing periodic motions of coiling ($\varphi$ varying almost linearly) and changing hemisphere ($\varphi$ $\approx$ constant).}
\label{figure_2}

\subsection*{Microtubule density tunes coiling dynamics and accessible trajectories}

We next examined if and how the surface density of microtubule bundles modulates the dynamics of the active layer. 

By varying the microtubule concentration during droplet formation, we generated cortices spanning a wide range of surface coverage, from dense, almost continuous nematic layers to sparse configurations containing extended voids (Fig. \ref{figure_3}a, top and Fig. S8 and S9, Movies S4 and S5). Across this entire range, the system robustly exhibits oscillatory dynamics described earlier, with repeated sequences of coiling within a hemisphere followed by crossings between hemispheres. However, while the global dynamical pattern is similar, the detailed trajectories and their variability depend sensitively on density.

At high surface coverage, the cortex forms a strongly coupled active nematic. Defects execute short, tightly constrained coiling trajectories around a pole, typically sweeping only a fraction of a full revolution before crossing to the opposite hemisphere (Fig. \ref{figure_3}a, bottom left). These trajectories are highly reproducible from cycle to cycle, as reflected by the narrow distribution of coiling angles (Fig. \ref{figure_3}b) and the tight clustering of paths in the ($\varphi,\theta$) plane (Fig. \ref{figure_3}c). On these time scales, we adopt an effective description of defects as self-propelled particles: a reduced description in terms of their positions - which is sufficient to account for their mutual repulsion - and their orientations - which determines the direction of active motion - suffices to examine the nearly deterministic oscillatory trajectories (Fig. \ref{figure_3}d).

Reducing the surface density progressively loosens the constraints that allow such a simple low-dimensional description. At intermediate densities, defects coil for longer angular distances within a single hemisphere, often completing more than half a turn before crossing (Fig. \ref{figure_3}a, bottom and 3b). At the same time, substantial cycle-to-cycle variability emerges: within the same droplet, some coiling events remain short while others extend over large azimuthal angles (Fig. \ref{figure_3}c). This increased spread reflects an expansion of the accessible configuration space, indicating that the mechanical coupling provided by the nematic field is weaker and that the system can explore a broader set of trajectories during each oscillation.

This increase in dynamical freedom is even more apparent when considering multiple oscillation cycles. At high density, defects almost always preserve their sense of rotation after crossing, leading to highly regular, unidirectional sequences over many cycles (Fig. \ref{figure_3}d and Fig. S10 and S11). In contrast, at lower densities, direction reversals become frequent, suggesting the coexistence of several dynamical attractors. Reduced density thus allows the system not only to explore more variability within a single coiling phase, but also to switch between distinct long-term patterns across successive hemisphere crossings (Fig. \ref{figure_3}d and Fig. S12 and S18).

\begin{figure}[p]
    \centering
    \includegraphics[width=1\textwidth]{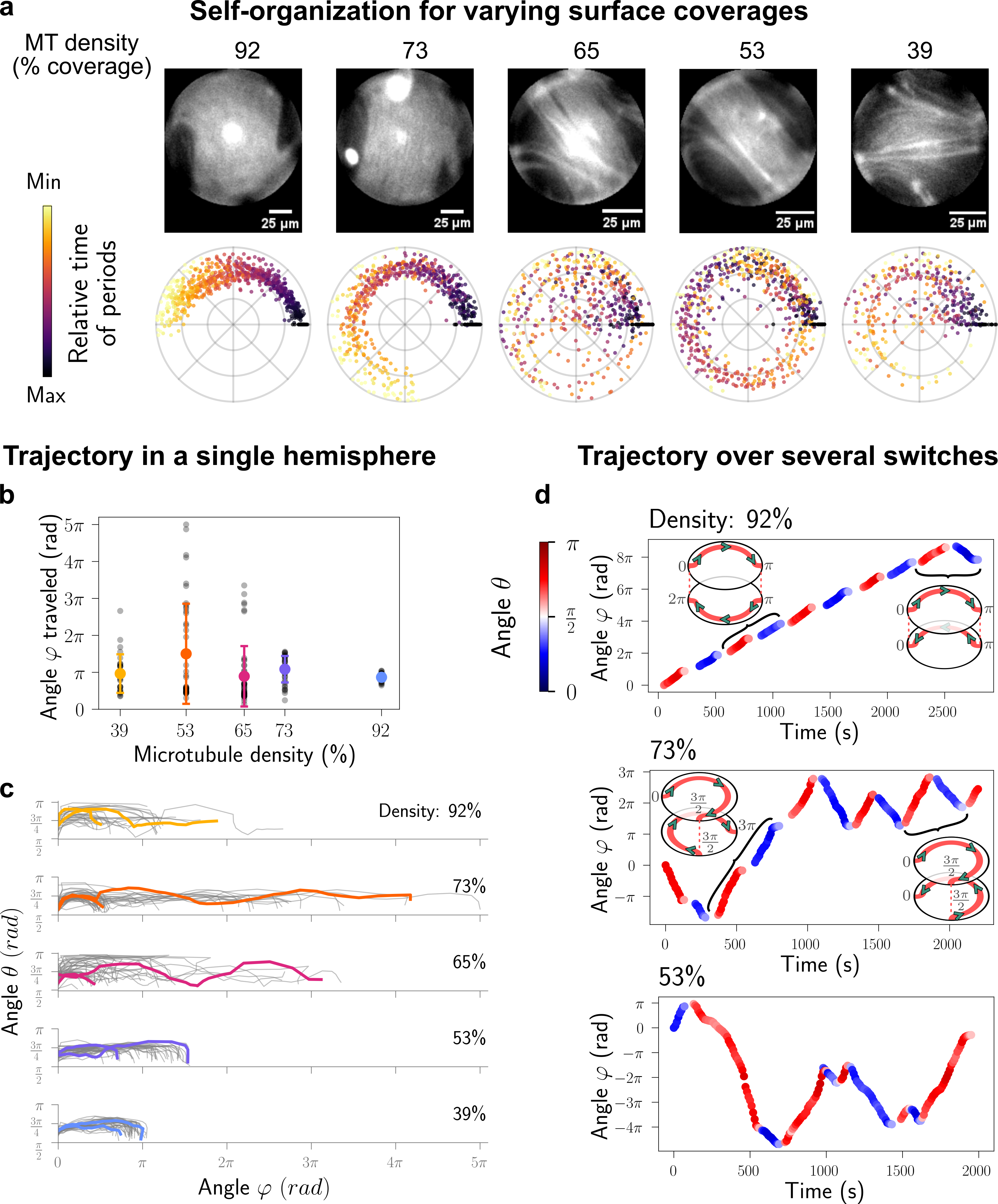}
\end{figure}

\clearpage
\captionof{figure}{\textbf{Density-dependent self-organization and trajectories of active nematics on a sphere.} \textbf{(a)} Representative fluorescence images of microtubule cortices confined to spherical droplets at decreasing surface coverage (92\%, 73\%, 65\%, 53\%, 39\%), together with polar plots showing the angular positions of +1/2 defects over one oscillation period. Trajectories are set to the same starting point and same direction for visualization. Color encodes relative time within the cycle. \textbf{(b)} Total azimuthal angle traveled during a single coiling phase within one hemisphere as a function of microtubule surface density. Colored symbols indicate mean values with standard deviation; gray symbols correspond to individual cycles. \textbf{(c)} Trajectories plotted in the ($\varphi,\theta$) plane for different densities. Thin gray lines show individual cycles, while colored curves indicate the 1st and 9th decile of trajectories. Increasing density leads to tighter, more reproducible paths, whereas reducing density results in broader exploration of configuration space. \textbf{(d)} Time evolution of the azimuthal angle $\varphi$(t) over multiple oscillation cycles for representative defects of droplets at different densities, highlighting periodicity and direction reversals. Insets illustrate schematic defect trajectories during coiling and crossing phases.}
\label{figure_3}

\subsection*{From a Nematic to a Filament-Based Description}

\begin{figure}[htbp]
    \centering
    \includegraphics[width=\textwidth]{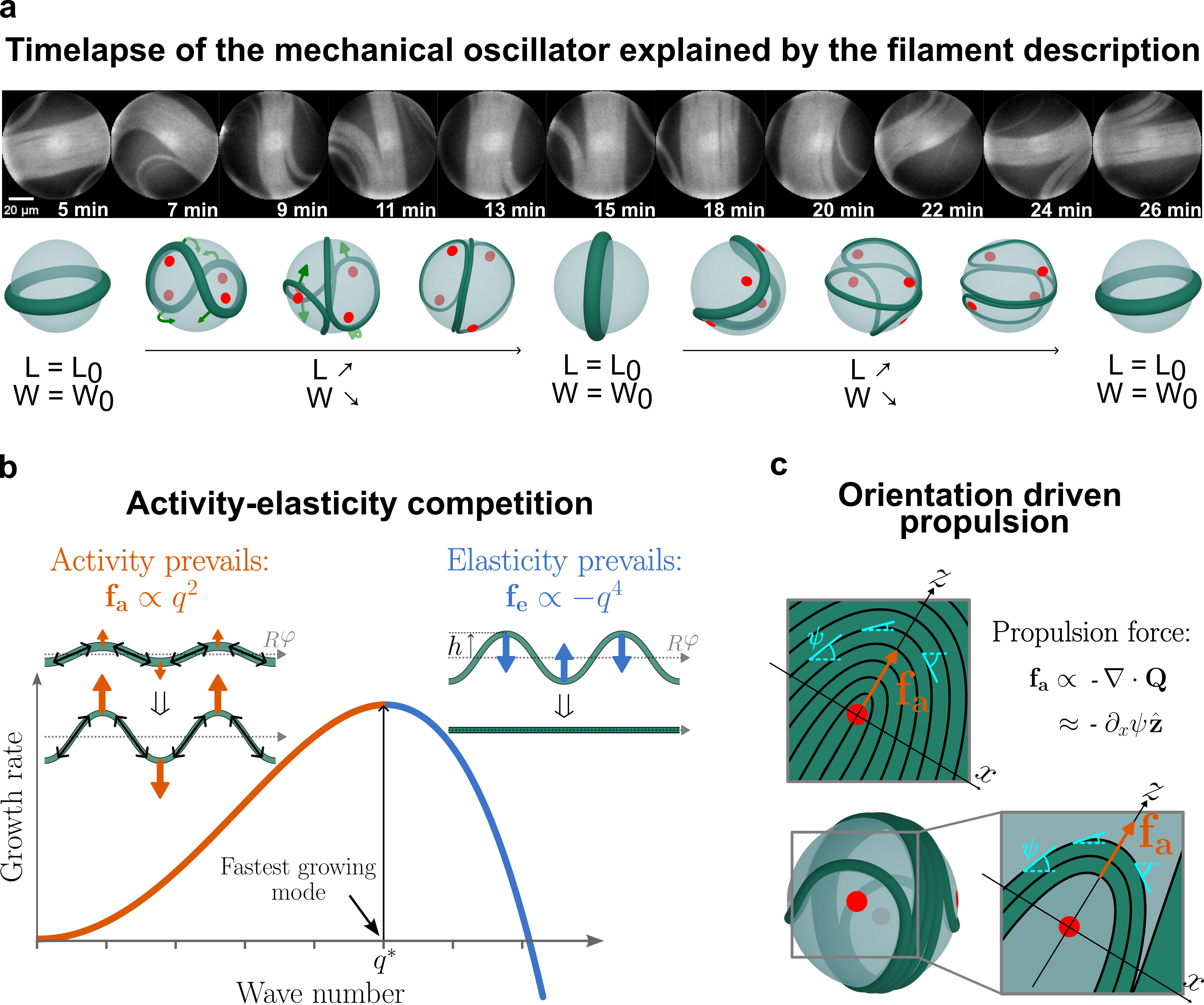}
    \caption{\textbf{Dynamics of an active filament.} \textbf{(a)} Time-lapse fluorescence images (top, 45\% MT density) and schematic filament-based interpretation (bottom) of one oscillation cycle at low density. The microtubule cortex behaves as a continuous extensile filament whose length $L$ and width $W$ evolve cyclically, providing an alternative description of the dynamics when nematic order weakens. \textbf{(b)} Growth rate of the filament length as a function of the wave number showing that two modes prevail at different values. For small wave numbers, the extensile activity prevails and the filaments extends. For high wave numbers, the elasticity prevails and the filament contracts. An optimum appears for a fastest growing mode q*. \textbf{(c)} The propagation of the extending filament is similar to the orientation-driven propulsion force from a nematic perspective.}
    \label{figure_4}
\end{figure}

At low densities, typically below 50\%, the nematic description becomes increasingly hard to justify. Rather than a uniform director field punctuated by isolated defects, the active layer organizes into a single thick extensile band that circulates along the spherical interface (Fig. \ref{figure_4}a). In this filamentous view, the four +1/2 ``defects'' correspond to regions of maximal curvature, two hills and two valleys, arising from the simplest buckling mode of an extensile filament confined to a sphere. Notably, when density is sufficiently low to permit extended voids, topological defects are no longer strictly required by orientational order because the orientation field is now defined only in a space that has the topology of a cylinder; nevertheless, the four-defect configuration persists, indicating that it is dynamically selected by the instability of the filament rather than imposed solely by topology.

This filament perspective also provides insight into the non-monotonic dependence of the coiling phase trajectories on density (Fig. \ref{figure_3}). At high density, the equatorial band occupies most of the available surface area, leaving little material in the polar regions to exert restoring forces; as a result, coiling remains short. At intermediate density, coiling matter in the hemispheres stabilizes the equatorial band against immediate buckling, leading to extended coiling. At very low density, however, there is insufficient material overall to counteract the filament’s bending instability, and coiling again becomes limited. We hypothesize that the balance between material stored in the band and in the surrounding hemispheres thus naturally explains the observed non-monotonic behavior.

Taken together, these results demonstrate that density does not induce a sharp transition between nematic and filament regimes, which one might expect to be very different from each other, but instead continuously tunes the effective degrees of freedom available to the system. Defect- and filament-based descriptions therefore offer complementary perspectives for a common underlying dynamics, which robustly generates periodic self-organization through the interplay of activity, curvature, and confinement.

\newpage
\subsection*{Theory of active nematics at varying areal coverage}
The density-dependent crossover suggests that the same organization can be viewed either as a defect dynamics or as an instability of an active band. At high areal coverage, the nematic director is defined over the full spherical surface. The Poincaré–Hopf theorem \cite{nelson_defects_2002,lubensky_orientational_1992, turner_vortices_2010, bowick_two-dimensional_2009,kamien_geometry_2002} then imposes a total topological charge $+2$, which is typically realized by four $+1/2$ defects \cite{chaikin_principles_1995}. Spatial variations of the director field around these defects generate active forces  \cite{marchetti_hydrodynamics_2013, aditi_simha_hydrodynamic_2002, voituriez_spontaneous_2005}, giving $+1/2$ defects their self-propelled character \cite{narayan_long-lived_2007, giomi_defect_2013}.

At lower coverage, the active material no longer forms a continuous nematic over the whole sphere. In this limit, topological arguments alone do not require the presence of defects. In the absence of activity, one would instead expect a stable defect-free equatorial band. The persistence of four defect-like bending regions therefore points to a dynamical selection mechanism (Fig \ref{figure_4}a and Movie S6).
We model this regime as a narrow active nematic band located near the equator and consider small transverse deformations 
$h(\varphi)$ of its centerline (see Supplementary section 11). A local tilt of the nematic director can be written as $\psi(\varphi)\approx\frac{1}{R}\partial_{\varphi}h$. The standard active force density, $\mathbf{f_a}=-\zeta\nabla\cdot\mathbf{Q}$, then gives, for a thin band, 
$\mathbf{f_a}=-\frac{\zeta}{R}\delta(z)\mathbf{\hat{z}}\partial_{\varphi}\psi \approx - \frac{\zeta}{R^2}\delta(z)\mathbf{\hat{z}}\partial_{\varphi}^2h$ where $\zeta$ is the activity coefficient, $\mathbf{Q}$ the nematic order tensor and $R$ the droplet radius.
This extensile active force amplifies curved regions of the band and therefore acts as an effective negative line tension \cite{senoussi_tunable_2019}. In Fourier space, the corresponding destabilizing contribution to the growth rate scales as $\zeta q_{\varphi}^2$. It is opposed by bending elasticity, which stabilizes short-wavelength deformations and contributes a term scaling as $-Kq_{\varphi}^4$, where $K$ the Frank elastic constant. Their competition selects a fastest-growing mode $|q_\varphi^*|=R \sqrt{\frac{\zeta}{2K\mu}}$, corresponding to a finite number of hills and valleys around the band (Fig \ref{figure_4}b and Fig. S19). Squeezing the droplets provides a direct way to test this prediction, because it increases the effective equatorial radius.

This effective one-dimensional description of a nematic band is very different and applies in a very different regime from a sphere on which the director field is defined almost everywhere (except at discrete points). However, the active force that deforms and destabilizes an equatorial band is the same one that drives the motion of $+1/2$ defects.


\newpage
\subsection*{Squeezing selects higher-order modes and reorganizes defect states}
To test how shape controls active self-organization, we mechanically deformed droplets by compressing them between the top and bottom surfaces of imaging chambers of varying height (Fig. \ref{figure_5}a, Fig. S20 and Movie S7). Squeezing transforms initially spherical droplets into oblate, pancake-like shapes, flattening the top and bottom regions while concentrating curvature along the equatorial side wall. As deformation increases, the active cortex forms a reproducible sequence of patterns: eye, triangle, square, and higher-order polygonal states (Fig. \ref{figure_5}a and Fig. S21 to S23 and S28). In these states, the microtubule-rich band folds around the equatorial region, as shown in both fluorescence images and schematics.

This sequence directly tests the band-instability theory introduced above. In the eye state, the band adopts the lowest non-trivial mode, $q=2$, corresponding to two hills and two valleys (Fig. \ref{figure_5}b). Increasing the effective equatorial radius selects higher modes, with triangle and square organizations corresponding to $q=3$ and $q=4$, respectively (Fig. \ref{figure_5}b). Quantifying the selected mode as a function of droplet radius confirms this trend: the mode number increases with radius, as predicted for active band undulations (Fig. \ref{figure_5}c).

Squeezing also reorganizes the topological charge distribution (Fig. \ref{figure_5}b). In the eye state observed at mild deformation ($\sim 25\%$ squeezing), four $+1/2$ defects satisfy the total charge
$+2$. At intermediate deformation ($\sim 33\%$ squeezing), the triangle state contains six $+1/2$ defects and two compensating $-1/2$ defects. At stronger deformation ($\sim 55\%$ squeezing), the square state contains eight $+1/2$ defects and two effective $-1$ defects. Thus, deformation does not change the global topological constraint, but redistributes charge through the nucleation of compensating positive and negative defects.

We will show that the position of these defects is selected by geometry. Squeezing creates a curvature inhomogeneity that generates a geometric potential acting with opposite signs on positive and negative defects (Fig. \ref{figure_5}d and Supplementary section 13). As a result, $+1/2$ defects are confined to the equatorial crown, while negative defects localize near the flattened top and bottom regions (Fig. \ref{figure_5}d). This spatial separation stabilizes mixed-charge configurations by limiting annihilation between opposite charges.

We next asked whether this geometric interaction is sufficient to account for the marked change in four-defect trajectories induced by weak squeezing. We revisited the defect-particle model introduced by Keber \textit{et al.} \cite{keber_topology_2014} and corrected the parallel transport of defect polarities between tangent planes on the sphere (Supplementary sections 14 and 15). This correction is required to recover the experimentally observed hemisphere exchanges and accurately reproduces the trajectories measured on perfectly spherical droplets (Fig.\ref{figure_5}e, top). We then incorporated the curvature-dependent geometric potential introduced in Fig.~5d. A weak oblate deformation is sufficient to reduce the azimuthal excursion of each defect from approximately $\pi$ to $\pi/2$, reproducing the trajectories observed experimentally in the eye state (Fig.\ref{figure_5}e, bottom). 

The main merit of this corrected particle model is to capture both the spherical and weakly squeezed four-defect states, and thus to highlight the contribution of geometry and non uniform curvature to the dynamics. 
However, this description is known to only effectively capture defects orientational dynamics \cite{shankar_topological_2022, Tang2019}. In fact, our numerical exploration of the dynamics in the case of  triangle and square states with respectively 6 or 8 (+1/2) defects, while it qualitatively reproduces the localisation of the +1/2 defects in the curved region, fails to quantitatively capture the seemingly periodic dynamics that are observed experimentally. Indeed, it is expected than by increasing the number of defects, the resulting deterministic trajectories will become more and more sensitive to refine details of the microscopic dynamics; we thus turned to a more generic and global classification of the defects dynamics, based on symmetry arguments and topological considerations, as explained below.

\begin{figure}[p]
    \centering
    \includegraphics[width=1\textwidth]{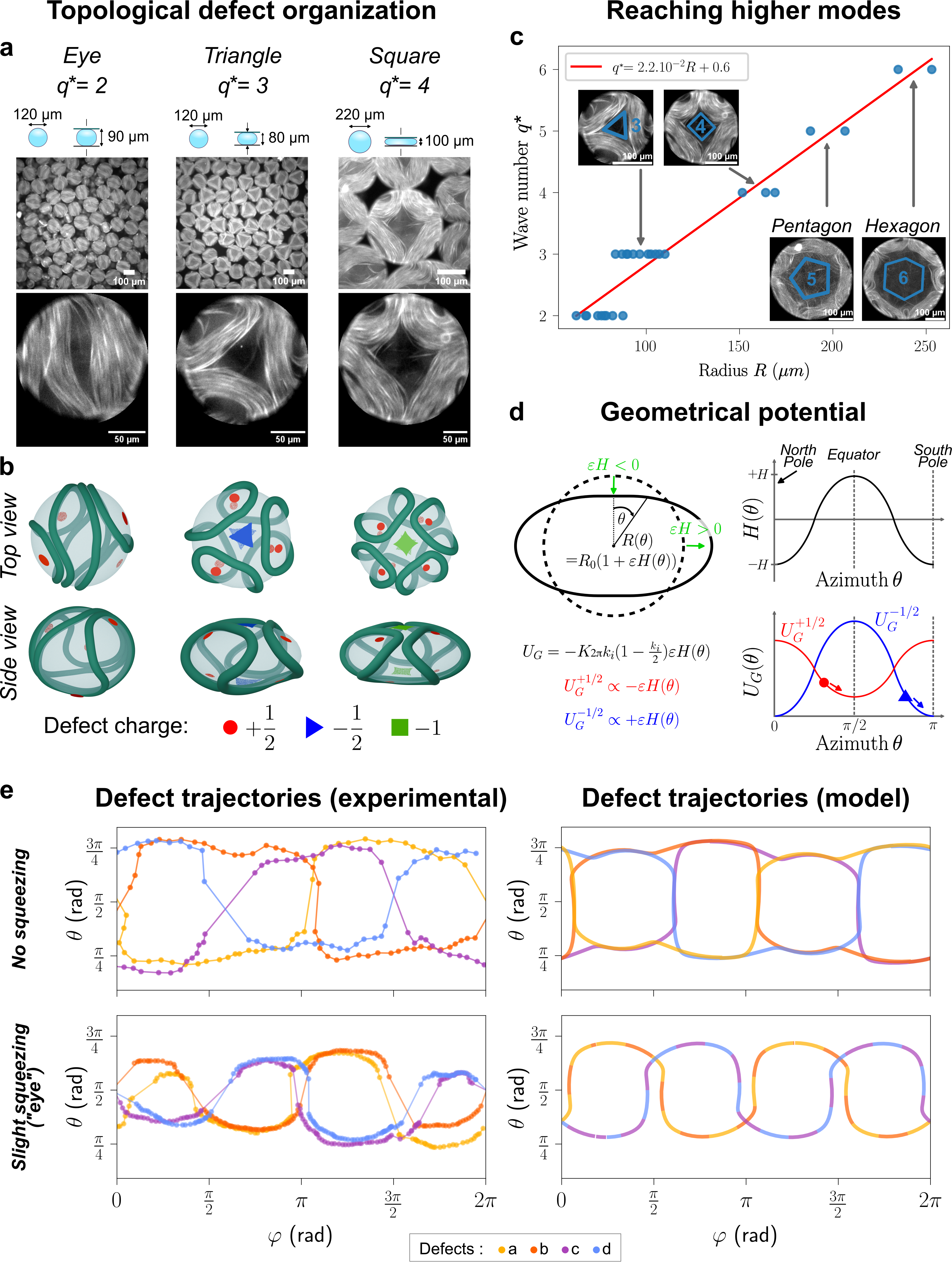}
\end{figure}

\clearpage
\captionof{figure}{\textbf{Curvature redistribution controls defect nucleation and organization under geometric deformation.} \textbf{(a)} Microscopy images of active nematic cortices confined to droplets subjected to increasing vertical compression (“squeezing”), transforming spherical droplets into oblate shapes. Mild squeezing produces an eye-shaped pattern, stronger squeezing yields triangular and square organizations. Schematics indicate droplet dimensions. \textbf{(b)} Schematic filament representations of defect organization for the eye, triangle, and square patterns. Red, blue, and green symbols denote +1/2, -1/2, and -1 defects, respectively. Top and side views illustrate how defects are confined to curved regions and migrate vertically along the droplet sides. \textbf{(c)} The wave number q of the folding filament evolves as a linear function of the radius of droplets squeezed in a 100 \textmu m high chamber. \textbf{(d)} Curvature inhomogeneity induced by compression gives rise to a geometric potential. This potential drives -1/2 defects toward the poles, where curvature is reduced, while +1/2 defects localize near the equator. \textbf{(e)} Experimental (left) and modelled (right) trajectories of the four +1/2 defects in the ($\varphi,\theta$) plane. Top, unsqueezed spherical droplet; bottom, weakly squeezed eye state. Colours identify the four defects. The model uses the corrected parallel transport of defect polarity and the geometric potential shown in d.}
\label{figure_5}

\subsection*{Nematic-to-hydrodynamic topology organizes defect trajectories}

Having shown that squeezing selects higher-order modes and spatially separates positive and negative defects, we next asked how this reorganized defect landscape constrains dynamics. We first quantified the trajectory of individual $+1/2$ defects in the different geometries (Fig. \ref{figure_6}a and Fig. S29). On the sphere, defects undergo large coiling trajectories and hemisphere crossings, exploring angular sectors close to $\pi$. In contrast, even mild squeezing strongly reduces the explored angle: in the eye state, individual defects move over sectors of approximately $\pi/2$. Stronger squeezing further subdivides the motion into sectors of approximately $\pi/3$ in the triangle state and $\pi/4$ in the square state (Fig. \ref{figure_6}a). Thus, squeezing transforms the non-stationary hemisphere-crossing dynamics of the sphere into serpentine trajectories confined to an equatorial ring.

Defects-as-particles models capture important aspects of the four-defect dynamics, but for six or eight (+1/2) defects, the complexity of the interactions cannot be captured.

We therefore use a complementary geometric viewpoint: rather than predicting the full flow field from microscopic interactions, we ask which hydrodynamic organizations are allowed by the topology and symmetry of the nematic texture. In active systems, a non-uniform configuration of the nematic orientation implies the existence of flow fields \cite{Serra2023,Head2024}.
The starting point is that nematic and flow fields have different symmetries since the former is a rank-2 symmetric, traceless tensor and the latter is a vector (Fig. \ref{figure_6}b). The nematic director is thus apolar and can host half-integer defects, whereas the velocity field is polar and can only host integer defects. In addition, both fields are not independent because in the local active-force approximation, the flow is assumed to be completely determined by gradients of the nematic order tensor, so that $\mathbf{v} \propto -\zeta \nabla \cdot \mathbf{Q}$. This force relation is shown in Supplementary section 12 (see also \cite{Khoromskaia_2017}) to generate a symmetry relation between the topological charges of the flow and the nematic fields $k_H = 2 k_N -1$ . This relation describes the local topology of the active forcing, not the full nonlocal hydrodynamic solution. 

This mapping gives a simple dictionary between nematic defects and hydrodynamic structures (Fig. \ref{figure_6}c). A $+1/2$ defect maps to $k_H=0$, corresponding to a polar pump-like structure with two counter rotating vortices, which is consistent with its self-propelled character. By contrast, negative nematic defects impose hydrodynamic hyper-saddles: a $-1/2$ defect maps to ($k_H=-2$), with three incoming and three outgoing branches, whereas an effective $-1$ defect maps to $k_H=-3$, with four incoming and four outgoing branches (Fig. \ref{figure_6}c). These branches act as separatrices that partition the surface into flow sectors. Importantly, this relationship between $k_H$ and $k_N$ holds only locally around a defect (because it is derived by assuming a single defect in infinite space). In order to capture the full pattern in the geometry of interest of a closed surface, one needs in addition to enforce the topological constraints for both the nematic field and the hydrodynamic field. 

We then used this dictionary to explore by inspection typical flow and nematic patterns that fulfil both the symmetry relation given above and the Poincar\'e-Hopf topological constraint of total charge $+2$ on sphere for both fields. The resulting states with increasing complexity are depicted in figure \ref{figure_6}d.


Direction-resolved trajectories of all $+1/2$ defects support this hydrodynamic-topological picture (Fig. \ref{figure_6}e and Fig. S30). In squeezed droplets, defects repeatedly revisit the same regions with consistent circulation directions, providing a proxy for the flow field (Fig. S31). Note however that the observed patterns, while consistent with our generic classification, are dynamic. The eye, triangle, and square patterns define hydrodynamic flows that guide the motion of the active defects. In this view, geometry and topology impose strong constraints on the nematic and flow fields that significantly restricts the range of accessible states, and allows for a systematic classification.

\begin{figure}[p]
    \centering
    \includegraphics[width=1\textwidth]{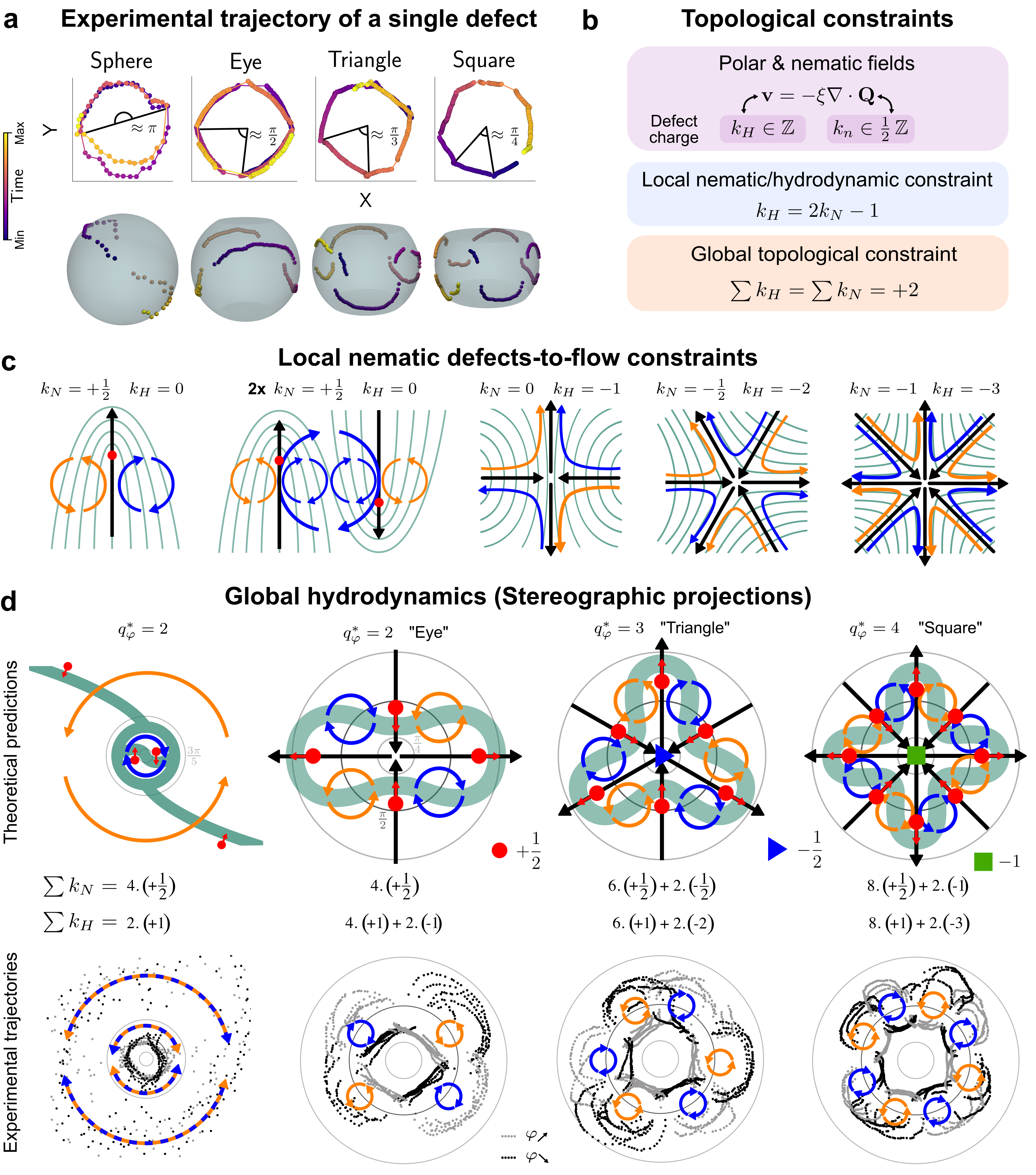}
\end{figure}

\clearpage
\captionof{figure}{\textbf{Defect trajectories and hydrodynamic topology in squeezed droplets.} 
    \textbf{(a)} Experimental trajectory of a single (+1/2) defect in a sphere and in squeezed droplets displaying eye, triangle, and square organizations. Top: projection of the trajectory in the equatorial plane. The color scale indicates time. Black sectors indicate the angular range explored by the defect. Bottom: corresponding three-dimensional trajectories displayed on the droplet surface. 
    \textbf{(b)} Local and global topological constraints. The local relation $k_H = 2 k_N -1$ gives the winding of the active-forcing field around an isolated nematic defect. The global +2 constraint follows from the Poincaré–Hopf theorem for a vector field on a closed spherical surface. 
    \textbf{(c)} Elementary nematic defect-to-flow mappings. Green lines represent the nematic field, arrows depict the associated local velocity field. From left to right: a (+1/2) nematic defect correspond to a pair of vortices turning in opposite directions. Two opposed +1/2 defects generate adjacent co-rotating cells, which are merged in the schematic representation. A charge-free splay region can generate a two-branch hydrodynamic saddle ($k_H=-1$). (-1/2) and (-1) nematic defects generate three- and four-branch hyperbolic structures ($k_H=-2$) and ($k_H=-3$). 
    \textbf{(d)} Stereographic projections viewed from the north pole of the schematic collective hydrodynamic organizations associated with the observed modes. Green bands represent the aligned microtubule-rich nematic band. Red and blue arrows indicate counter-rotating flow cells, and black arrows show the separatrices or imposed by the hydrodynamic defects. In squeezed droplets, $q_\varphi^*=2$, $q_\varphi^*=3$, $q_\varphi^*=4$, correspond to two, three, and four hills and valleys of the nematic band, which are associated with four, six, and eight hydrodynamic vortices, respectively. 
    \textbf{(e)} Experimental trajectories of all detected (+1/2) defects shown as stereographic projections viewed from the north pole, for the same sequence of geometries. Black and gray points indicate trajectory segments moving in opposite circulation directions. These direction-resolved trajectories are used to infer the red and blue counter-rotating vortices shown in \textbf{(d)}.}
\label{figure_6}

\subsection*{Discussion}
Active nematics on spheres have provided one of the clearest examples of how topology can organize active matter. In the seminal experiment of Keber \textit{et al}. \cite{keber_topology_2014}, spherical topology imposes four $+1/2$ defects, whose self-propulsion drives oscillations between tetrahedral and planar configurations. 


Our droplets recover this canonical four-defect dynamics, but also show where this description reaches its limits. The Keber picture, and the theoretical models that followed it \cite{henkes_dynamical_2018, zhang_dynamics_2020, alaimo_curvature_2017} are most natural when the active layer is a continuous nematic coating a surface of nearly uniform curvature. Here, both assumptions can be relaxed: density changes the continuity of the active material, and squeezing redistributes curvature. The relevant degrees of freedom then cease to be only the positions and polarities of four (+1/2) defects.

This first appears when surface density is reduced. At high coverage, the system is well described as an active nematic whose defects are required by the Poincar\'e–Hopf constraint. At lower coverage, however, the active material no longer covers the full surface so a total defect charge of $+2$ is not a topological requirement. Yet a four-fold organization persists. This persistence suggests that the observed structures are selected not only by topology, but also by a mechanical instability of the active material. In this regime, the cortex is better viewed as an extensile band whose bending produces defect-like regions. The density series therefore connects two descriptions that are often treated separately: a defect-based active nematic and a filament-like active band. The band-instability theory gives a common mechanism for this continuity, because the same active forcing that propels $+1/2$ defects can act as a negative line tension that destabilizes a curved band, while bending elasticity selects a finite mode.

Squeezing provides a direct way to explore active matter on more complex curved surfaces, beyond the homogeneous curvature of a sphere. By deforming the droplets, we experimentally reshape the curvature landscape while keeping the active material unchanged, allowing us to probe how spatial curvature heterogeneity selects the organization of the active layer. This connects our results to broader studies of active nematics on curved surfaces \cite{Pearce2019}, including spheres, ellipsoids \cite{clairand_dynamics_2024} and toroids \cite{ellis_curvature-induced_2018,Nambisan2024}, where curvature and topology have been shown to control defect localization, unbinding and motion. Increasing the effective equatorial radius selects higher band modes, as predicted by the theory. Additional $+1/2$ defects are compensated by negative defects so that the total charge remains $+2$. An important point is that curvature heterogeneity spatially separates opposite charges, stabilizing mixed-charge configurations that would otherwise tend to annihilate. Geometry therefore acts not merely as a boundary condition, but as a selector and spatial organizer of accessible topological states.

Despite the complexity arising from the large number of interacting defects, the system exhibits remarkably regular and persistent dynamics. The intrinsic coupling between the nematic field and the polar flow field, together with global topological and geometrical constraints, defines selection rules for robust dynamical states of increasing complexity. This reduced description yields flow architectures that are consistent with the experimentally observed trajectories, showing how the organization of the nematic field can constrain and predict the resulting flow dynamics.



This approach resonates with observations in living systems, where geometry and topology are not merely passive constraints \cite{Tan2024}. Curvature gradients can bias tissue-scale cytoskeletal flows \cite{gehrels_curvature_2023}, as shown during Drosophila gastrulation, and topological defects in actin or cellular nematics can act as mechanical organizers of morphogenesis, as in Hydra regeneration \cite{maroudas-sacks_topological_2021,Ravichandran2025} or confined cellular monolayers \cite{guillamat_integer_2022}. Our droplets do not reproduce these systems and deliberately remove biochemical regulation. Their relevance is instead reductionist: they show that activity, density and curvature alone can generate persistent cycles, defect nucleation, charge separation and flow-guided trajectories. In this sense, they isolate a physical layer of morphogenesis on which biochemical feedback could act in living matter.

Future work could explore dynamically evolving geometries, as well as surfaces with different topological charge constraints and regions of negative curvature, together with the theoretical framework needed to describe these new configurations. Such directions could bring us closer to understanding morphogenesis, moving from geometry as a static organizer toward the more biological situation in which mechanics, activity and shape continuously reshape one another.

\newpage
\subsection*{Materials and methods}
\subsubsection*{Protein preparation}
The motor protein used in this work was K401, a truncated kinesin-1 construct containing the first 401 residues of Drosophila melanogaster kinesin \cite{subramanian_two_2007} and bearing a biotin tag, allowing multimeric motor clusters to form in the presence of streptavidin \cite{nedelec_self-organization_1997}. Kinesins were expressed in E. coli and purified as described previously \cite{senoussi_tunable_2019}. Tubulin and rhodamine-labelled tubulin were purchased from Cytoskeleton, Inc. and dissolved at 10 mg.mL$^{-1}$ in tubulin buffer containing 80 mM PIPES, 2 mM MgCl$_2$, and 0.5 mM EGTA. GMPCPP-stabilized microtubules were polymerized in 1X PEM-KOH buffer, 10\% glycerol, 0.2 mM DTT, 0.6 mM GMPCPP, and 5 mg.mL$^{-1}$ tubulin containing 2.5\% rhodamine-labelled tubulin. The mixture was incubated for 30 min at 37 °C and then for 6 h at 25 °C. Polymerized microtubules were flash-frozen and stored at -80°C until use.

\subsubsection*{Active material composition}
The active material was assembled from GMPCPP-stabilized microtubules, K401 kinesin motors, streptavidin, Pluronic F-127, and an ATP-regeneration system. After mixing inside droplets, the standard active mixture contained 1X PEM-KOH buffer, 3 mM MgSO$_4$, 2\% (w/v) Pluronic F-127 (Sigma-Aldrich), 3 mM DTT, 150 µg.mL$^{-1}$ glucose oxidase, 20 mM D-glucose, 25 µg mL$^{-1}$ catalase, 1 mM Trolox, 0.5 mg.ml$^{-1}$ bovine serum albumin, 5 µg.mL$^{-1}$ creatine kinase, 10 mM creatine phosphate, and 300 µM NPE-caged ATP (adenosine 5'-triphosphate, p3-(1-(2-nitrophenyl)ethyl) ester, disodium salt) (Thermo Fisher Scientific). Unless otherwise stated, droplets contained 75 nM kinesin motors, 75 nM streptavidin, and 1 mg.mL$^{-1}$ microtubules after mixing. For the density-dependent experiments, the amount of microtubules and kinesins encapsulated in droplets was varied by changing the relative injection pressures while keeping the motor-to-microtubule ratio fixed. The compositions used for the main figures are reported in Supplementary Table S1.

\subsubsection*{Microfluidic droplet generation}
Water-in-oil droplets were produced using a PDMS microfluidic device with a flow-focusing geometry, as described in the supplementary materials. The device contained three aqueous inlets and one oil inlet. The aqueous streams were separated before droplet formation to prevent premature contact between microtubules and kinesin clusters, which otherwise promotes network formation and can clog the device. One aqueous stream contained microtubules, one contained buffer, and one contained kinesin motors and streptavidin. The three aqueous streams merged immediately before intersection with the oil phase, allowing the active mixture to form only inside the droplets. The oil phase consisted of fluorinated oil Fluo-Oil 7500 with 2\% Fluosurf-C surfactant (Emulseo). Flow was controlled using a pressure-based flow controller (LineUp Flow EZ™, Fluigent), and droplet formation was monitored by bright-field and fluorescence microscopy (ZEISS Axio Zoom.V16). The resulting emulsions were collected from the chip outlet in pipette tips.

\subsubsection*{Droplet activation and confinement}
Before activation, ATP was kept in a caged form to prevent motor activity during droplet generation. Droplets were transferred into 1 cm x 1 cm silicon chamber \cite{lobato-dauzier_silicon_2023} of controlled height and sealed with a coverslip to form a monolayer. The silicon chambers were built in the cleanroom of the FEMTO-ST lab (Besançon). The coverslips were treated with a hydrophobic surface treatment (ST-2, Emulseo). The chamber height was chosen according to the droplet size and the desired degree of confinement. For unconstrained spherical droplets, chamber heights were chosen to avoid significant deformation. For squeezed droplets, droplets were compressed between the top and bottom surfaces of chambers of reduced height, producing oblate shapes with flattened polar regions and a curved equatorial side wall, as described in the supplementary materials. ATP was uncaged by illuminating the filled chambers with a 365 nm UV LED (Wurth Elektronik), hence activating motor activity. After activation, droplets were left to reorganize until microtubules accumulated at the oil–water interface and formed an active cortical layer.

\subsubsection*{Fluorescence imaging}
Droplets were imaged using a ZEISS Axio Zoom.V16 epifluorescence macroscope equipped with an HXP 200C fluorescence source, a CMOS camera (ORCA-Flash4.0, Hamamatsu), and a motorized stage controlled with MicroManager \cite{edelstein_advanced_2014}. Rhodamine-labelled microtubules were imaged in the red channel. A fluorescent dye added to the kinesin-containing stream was used to monitor relative motor concentration across droplets. Samples were maintained on a temperature-controlled stage to limit heating during imaging and prevent microtubule depolymerization. The large depth of field of the macroscope allowed nearly complete visualization of either the top or bottom hemisphere of each droplet. For dynamical measurements, top and bottom hemispheres were imaged sequentially or in paired acquisitions, allowing reconstruction of the active cortex over the full droplet surface.

\subsubsection*{Surface density estimation}
The surface density of microtubules was defined as the fraction of the droplet surface covered by microtubule-rich material after interfacial accumulation. At high density, this quantity was estimated directly from fluorescence images by segmenting microtubule-rich and microtubule-poor regions and normalizing by the apparent droplet surface. For lower densities, where phase-separated domains were less sharply defined, the nominal surface density was inferred from the injected microtubule concentration and droplet size, using the controlled microfluidic composition and the volume-to-surface ratio of the droplets.

\subsubsection*{Defect tracking and analysis}
Topological defects were identified from fluorescence images as microtubule-poor, tear-shaped regions associated with strong curvature of the surrounding nematic texture. The position of each (+1/2) defect was assigned to the head of the tear-shaped region, corresponding to the point of highest curvature in the local director field. At low surface density, where the nematic texture was less continuous and defects were less sharply defined, the same criterion was used to track defect-like bending regions. Defects were tracked manually in Fiji/ImageJ and analyzed using custom Python scripts. For spherical droplets, defect positions measured in microscopy images correspond to projections of points on the droplet surface into the imaging plane.

\subsubsection*{Reconstruction of missing equatorial segments}
Because defects are difficult to identify near the equator, short gaps appear in the spherical trajectories. These gaps were reconstructed using the symmetry and periodicity of the four-defect dynamics, as described in the supplementary materials section 4.

\subsubsection*{Analysis of squeezed droplets}
Squeezed droplets were analyzed from top and bottom fluorescence image sequences. Because the surface was no longer spherical, defects trajectories were analyzed using a Mercator projection, as described in the supplementary materials section 7.3 (Fig. S24 to S27).

\subsubsection*{Manuscript writing}
Generative AI tools were used to assist with translation, wording and English-language editing.

\newpage
\subsection*{Acknowledgments}
We thank Samuel Bell, Jérémy O'byrne, Guillaume Sarfati and Elie Wandersman for insightful discussions, Nicolas Fix-Boulier for technical support. This work has been funded by the French ANR LiliMat program (Grant No. ANR-22-CE06-0004, J.-C. G.).


\clearpage

\printbibliography


\end{document}